\documentclass{IEEEtran}
\IEEEoverridecommandlockouts
\usepackage{graphicx}
\usepackage{booktabs}
\usepackage{amsmath}
\usepackage{comment}
\usepackage[linesnumbered,ruled,vlined,boxed,commentsnumbered]{algorithm2e}
\usepackage{lscape}
\usepackage{tabularx}
\usepackage{longtable}
\usepackage{multirow}
\usepackage{subfigure}
\usepackage{tikz}
\usepackage{pgfplots}
\usepackage{pgf}
\usepackage{cite}
\usepackage{amsmath}
\allowdisplaybreaks[1]
\usetikzlibrary{patterns}

\usepackage[colorinlistoftodos]{todonotes}

\usepackage{color,soul}
\definecolor{myred}{RGB}{223,42,42}
\definecolor{myblue}{RGB}{30,144,255}
\definecolor{mygreen}{RGB}{88,212,88}
\definecolor{mypurple}{RGB}{138,43,226}

\normalsize

\begin{document}
\title{A Novel Initialization Method for Hybrid Underwater Optical Acoustic Networks}

\author{Yuanhao~Liu,
        Fen~Zhou, ~\IEEEmembership{Senior~Member,~IEEE,}
        Tao Shang
% \thanks{Yuanhao Liu is with the State Key Laboratory Integrated Service Networks, School of Telecommunications Engineering, Xidian University, China. He is also with CERI-LIA, Computer Science Laboratory, University of Avignon, France. (email: yuanhao\_liu@foxmail.com).}
\thanks{Fen Zhou is with IMT Nord Europe, Institut Mines-T\'el\'ecom, Univ. Lille, Center for Digital Systems, France. (email: fen.zhou@imt-nord-europe.fr).}
% \thanks{Tao Shang is with the State Key Laboratory of Integrated Service Networks, School of Telecommunications Engineering, Xidian University, China (email: shtsun\_sjtu@hotmail.com).}
% \thanks{Zuqing Zhu is with the School of Information Science and Technology, University of Science and Technology of China, China (email: zqzhu@ieee.org).}
% \thanks{Juan-Manuel Torres-Moreno is with CERI-LIA, University of Avignon, France.}
% \thanks{A preliminary version of this work has been accepted by IEEE Globecom conference 2020 \cite{LiuGlobecom2020}}
% \thanks{The work is partially supported by Eiffel Excellence Scholarship (No.P745849E), China Scholarship Council (202006960046), Campus France PHC Cai Yuanpei 2019 project (44016XA),the 863 High Technology Plan of China (2013AA013402), and the National Natural Science Foundation of China (61172080, 61771357). Xidian University, China.}
}
\maketitle

\begin{abstract}
%To satisfy the high data rate requirement and reliable transmission demands of underwater scenarios, the hybrid wireless optical and acoustic communications system is considered a valid solution.
To satisfy the high data rate requirement and reliable transmission demands in underwater scenarios, it is desirable to construct an efficient hybrid underwater optical acoustic network (UWOAN) architecture by considering the key features and critical needs of underwater terminals. In UWOANs, optical uplinks and acoustic downlinks are configured between underwater nodes (UWNs) and the base station (BS), where the optical beam transmits the high data rate traffic to the BS, while the acoustic waves carry the control information to realize the network management. In this paper, we focus on solving the network initializing problem in UWOANs, which is a challenging task due to the lack of GPS service and limited device payload in underwater environments. To this end, we leverage  acoustic waves for node localization and propose a novel network initialization method, which consists of UWN identification, discovery, localization, as well as decomposition. Numerical simulations are also conducted to verify the proposed initialization method. 
%Due to the limited payload and battery storage,  underwater nodes (UWNs) are only equipped with energy-efficient optical transceivers and the acoustic receiver. The asymmetrical links between the UWNs and the base station (BS) are then configured with optical uplinks and acoustic downlinks, where the optical beam transmits the high data rate traffic to the BS, and the acoustic waves carry the control information to realize the network management. In particular, we propose a novel method to address the challenging networking initialization problem in UWOANs, which  with the help of the underwater acoustic waves.onsisting of identification, discovery,location,  and  decomposition  algorithm  is  designed  in  deta %Then, we propose the solutions to several underwater optical wireless networking conundrums. Next, we explain the critical issues for the proposed UWOAN and analyze their impacts. We finally discuss the open challenges and future works that need to be further explored.
\end{abstract}

\begin{IEEEkeywords}
Hybrid optical acoustic underwater network, underwater optical communication, network initialization 
\end{IEEEkeywords}

\section{Introduction}
The Internet traffic has been growing rapidly over the past decade, at the rate of more than 30\% each year \cite{Zhu2013_JLT,Gong2014_JLT,YLiu2021TNSM}. Also with the increasing interest in ocean exploration, marine wireless data has shown the features of big data with the huge volume and high value \cite{YLi2018WC,Lu2015_Network}. Over 4,000 buoys and floats offsite link take daily measurements at the ocean surface as well as thousands of meters below. The data includes sea surface temperature, ocean chemistry, currents, sea level, sea ice, and heat content with its volume up to 96 PB in 2016 \cite{NOAA}. The underwater wireless transmission urgently demands high speed and robust communications. To build an ecological ocean exploration and monitoring, the underwater communications need to face the trade-offs among the energy consumption, data rate, network scale, payload of the underwater devices, etc \cite{ZZeng2017CST,YLiu2018ICTON}.

In underwater environments, the radio frequency (RF) signal is highly attenuated by saltwater. The bandwidth of the military RF communications is usually from 30 Hz to 300 Hz, which requires huge antennae with a length up to $100 \hspace{1mm} m$ for the submarines. On the other hand, optical communications have shown a strong capacity to carry the huge volume of data traffic, thanks to its high spectral efficiency and the great progress on hardware \cite{Gong2013_JOCN,Yin2013_JOCN}. Naturally, the underwater wireless optical communication (UWOC) is then considered as the promising communication mode to meet the requirement of the high bandwidth, low latency and low energy consumption in the underwater environments \cite{BShihada2020CM,ZZeng2017CST,ACelik2020CM,YLiu2018ICTON}. Several commercial optical transceivers have been put into the solutions of underwater data transmission. For instance, the transmission rate of UWOC is up to 5 Mbps with a distance of $200$ meters in clear water. However, the UWOC suffers from networking difficulties due to its narrow beam and line-of-sight (LOS) communication mode. Meanwhile, the conventional underwater acoustic wireless communication (UWAC) has been considered as the most developed underwater communication technique. Although the UWAC enables to benefit from the long link range and the broadcast channel, it also has the deficiencies, such as low bandwidth and low propagation speed ($1,500 \hspace{1mm} m/s$). Based on the pros and cons of UWOC and UWAV, two communications techniques can be integrated together to realize the underwater data transmission by carrying different data. The wide-angle and low-speed acoustic waves are suitable to transmit the control information for signaling and networking. The LOS optical beam is suitable for high-speed and low-latency data, such as real-time multimedia data and big-volume hydrographic data.

In underwater scenarios, underwater nodes (UWNs) are usually undersea devices with limited battery storage and payload, such as diving buoys, and underwater sensors, autonomous underwater vehicles (AUVs), and unmanned aerial vehicles (UAVs). They are expected to be widely deployed all over the ocean and to continuously collect data for the long term. However, the acoustic data transmission is too energy-expensive for them since the energy efficiency is about 100 bits per Joule. Meanwhile, the energy efficiency for UWOC is up to 30,000 bits for each Joule. As a consequence, the acoustic transmitter is not suitable to be equipped in the energy-sensitive UWNs, which require low power consumption to extend the active cycle. Nonetheless, UWAC can be implemented in the BS, which is usually equipped with effective energy replenishment. The most common BS is the boat, buoy, and seafloor-wired platform. Such an asymmetrical network architecture, namely the hybrid underwater optical acoustic networks (UWOAN), is suitable for most underwater applications with unidirectional high-speed transmission, such as marine observation, disaster warning, object detection, ocean exploration, etc.

\begin{figure*}[t]
  \centering
  \includegraphics[width=160mm]{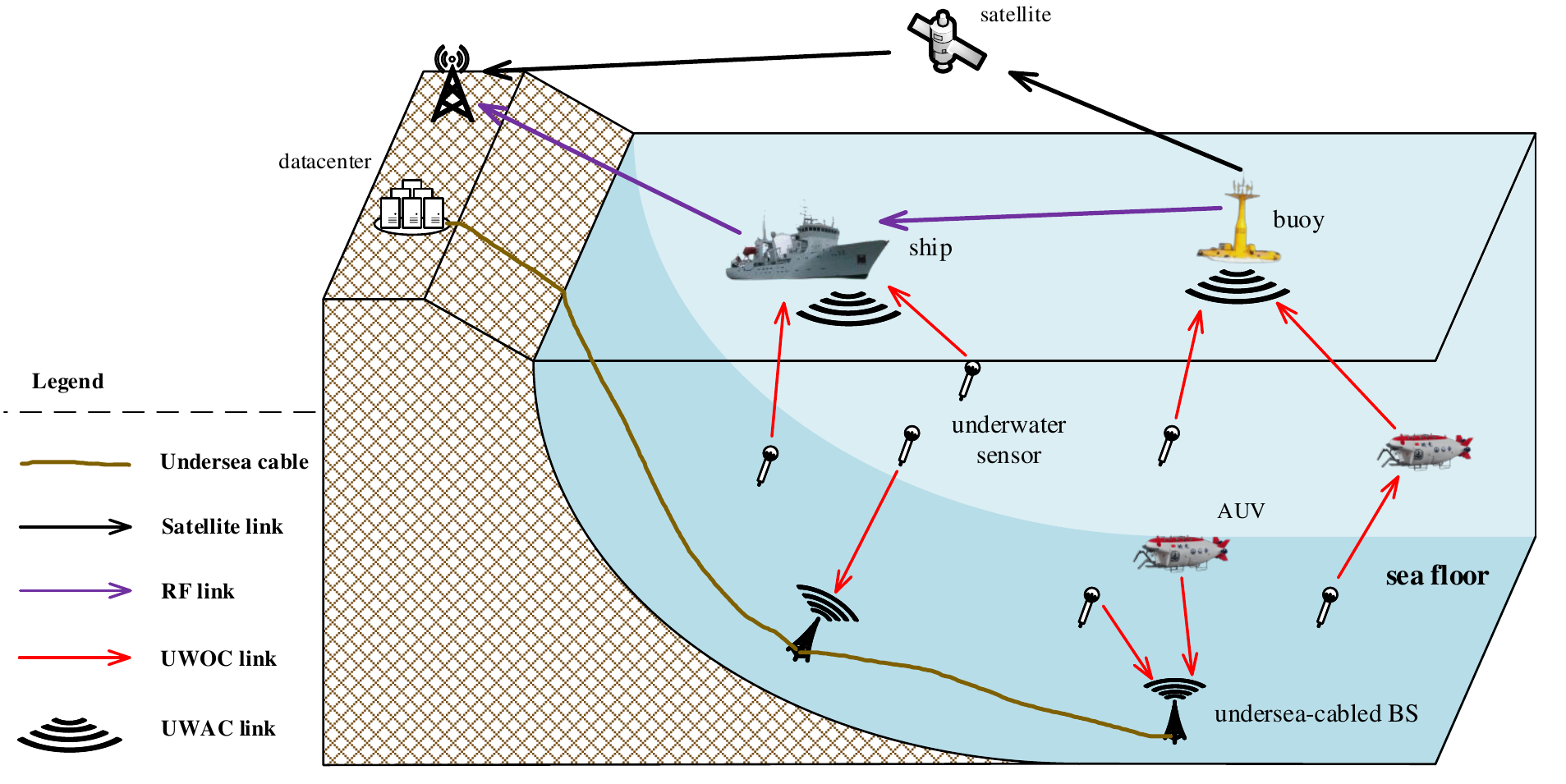}
    \vspace*{-0.5\baselineskip}
    \caption{Architecture of the UWOAN. Only optical uplink and acoustic downlink are implemented in the UWOAN.}
    \label{fig:Arch}
\end{figure*}

In the UWOAN construction stage, the UWOC uses laser, or high-power LED, as the transmitters with narrow beams, which demand strict conditions to send optical beam fallen on the optical receiving lens. The orientation information is essential for the UWNs to establish the optical links and then to route with the geographic coordinates. However, as the global positioning system (GPS) signal cannot propagate through seawater, one important question that arises for the networking initialization is how to economically and practically transfer the geographic information at the beginning. Usually, the UWNs need buoys, ships, or other reference nodes deployed above the water surface to locate themselves and the access point. Hence, it is a contradiction that the UWNs need to communicate with the BS to get the coordinates, yet the link-establishing is based on their coordinates. 

The existing works of hybrid optical and acoustic communications have focused on end-to-end performance, optical alignment, relay configurations, multiple access, routing protocol, etc. Most of them assumed the networking initialization is achieved with either the duplex acoustic communications or underwater optical communications with the wide-angle LED beam within 20 meters \cite{Saeed2019AH}. Such assumptions can get around the networking initialization problem of UWOC. However, these configurations are not practical for today's underwater terminals, for they are either bulky, energy-expensive, or transmission range limited, as aforementioned. The existing works have demonstrated the potential of hybrid optical and acoustic communications, yet, to the best of our knowledge, no work has been found that is exploring the initialization of the asymmetrical underwater network. Our work aims to fill this gap by introducing a practical UWOAN architecture, out of consideration for the eager needs on lightweight, low power consumption, low payload, and medium-range data transmission of the underwater terminals. The proposed UWOAN differs from the existing works as we encourage an economic underwater data transmission, where the asymmetrical links are configured with only optical uplinks and acoustic downlinks. We hope that the UWOAN can better serve the underwater applications rather than add to the already heavy burden of underwater terminals. The contributions of this paper are summarized as follows.

In this article, the implementation of the UWOAN is presented as asymmetrical with optical uplinks and acoustic downlinks, considering the realistic scenarios and features. A solution to the networking conundrum of the presented UWOAN is explored, and a simple and practical networking initialization method to build a UWOAN is proposed. Then the initialization method consisting of UWN identification, discovery, location, and decomposition is designed in detail. The simulations are also conducted with the proposed system. Numerical results show high practicality with different scenarios. At last, we conclude the article with a comprehensive discussion about future work.

\section{A Hybrid Underwater Optical Acoustic Network Design}

As the main approach for ocean exploration, the large number of underwater terminals should be widely distributed for the long term. The economic input for such human activity is huge. For instance, over 16,000 buoys and floats are deployed all over the world for the project of Array for Real-time Geostrophic Oceanographic (Argo). However, only less than 4,000 are working online \cite{ARGO}. The ocean environment is unfriendly to the monitoring and data collection terminals. The design of the high-powered but low-cost underwater terminal is still an urgent challenge to be solved. To this end, we focus on the network architecture that is beneficial for reducing the power consumption as much as possible for the UWNs. Basically, the UWNs only need to passively receive the acoustic signals and send the optical beam toward the informed emission directions, and the BS will handle the rest. The architecture of the UWOAN is shown in Figure \ref{fig:Arch}, in which the acoustic links are established as unidirectional. Note that the data transmission from BS to the on-shore datacenter is not a part of the UWOAN, \textit{i.e.} the wired transmission via subsea cable, the wireless communication to the on-shore access points, and the satellite transmission.

\subsection{UWN}
The UWNs are usually the free-drifting underwater terminals, which are limited by the low energy budget and payload, such as AUV, underwater buoys, underwater sensors, platform anchored undersea, etc. Thus, we aim to relieve the payload and the power requirement for the UWNs. Considering that the acoustic transmitters are not suitable to be equipped in the UWNs due to the bulky volume, and high power consumption, the UWNs in the UWOAN are only equipped with passive acoustic receivers and optical transceivers, which are cost-efficient for both volume and power. In other words, the UWNs can receive acoustic and optical signals and transmit optical beams, but they cannot send acoustic waves, which is the main difficulty in networking initialization. Furthermore, due to the trade-off between the UWOC transmission range and the optical beam width impacted by water absorption and scattering \cite{ZZeng2017CST}, the optical transmitters on UWNs are based on narrow-beam LED or lasers to expand the transmission range. The UWNs also need to be equipped with a high-accurate underwater digital compass to distinguish the horizontal direction, for which the reason is then discussed in Section \ref{sec:UWN-identification}.

\subsection{BS}
The BSs are buoys, ships, submarines, undersea-cabled platforms, etc., which have either energy harvest devices, \textit{e.g.} solar panels, or enough fuel supply. A BS is the center node of a cluster in the UWOAN. Its main works are helping to establish the optical links, collecting data from the UWNs, and transferring the data to satellites or on-shore data centers. Note that in such scenarios, the BS is not required to send an optical beam to the UWNs. However, it is fully capable to establish optical downlinks from the BS to the UWNs. The discussion of the optical downlink is not covered in this article for the lack of the corresponding applications. Furthermore, the BS is also equipped with a sonar system to detect the UWNs. The location information is then broadcasted through acoustic signals to instruct the link establishment.

\subsection{Link Configurations}
Based on the above system implementations, the signal transmission downward is a narrow-band acoustic downlink from the BS to the UWNs, where the wide-angle acoustic signal carries the control and coordinates information to build and maintain the optical links. The data transmission upward is the high-speed optical uplink from UWNs to BS, in which the optical signal is the highly directional narrow optical beam, which transmits the high-volume data such as video, picture, multimedia, etc.

\section{Network Initialization for UWOAN}

\subsection{UWN Discovery}
Lacking the GPS service underwater, the UWNs need to find another feasible means of locating the neighboring nodes and themselves. In free space optical communications (FSO), the pointing, acquisition, and tracking (PAT) mechanisms are used to establish and maintain the wireless optical links between satellites. Such mechanisms require servo motors of different precision, and complicated scan methods to adjust the orientation angle of the board beam. It is expensive and power-consuming for the UWNs. For the BS, it is suitable to discover and locate the UWNs by active sound echo with sonars. The specific acoustic frequency can be set as the trigger information to active the UWNs. The acoustic signals can also be broadcasted with the location information to the detected UWNs.

\subsection{UWN Localization}
\label{sec:location}
Once the BS enters the sea area of deployed UWNs, it starts to detect the underwater objects with a sonar system. The UWNs are then awakened by the specific acoustic waves, and the activated UWNs wait to receive the next acoustic signals. In this phase, the UWNs are considered as “muted”. They cannot respond to the BS, because the narrow optical beam transmission requires precise emission angles. After the UWNs discovery and location in the BS, the BS will allocate each detected UWNs the unique network ID with the downward acoustic channel resources, which can be the fixed time slots leveraging the time-division multiple access (TDMA). The BS provides the relative position information of each node within the allocated corresponding time slots, which are broadcasted through the wide-angle acoustic wave to cover the area.

Most UWNs are passively drifting due to underwater currents. For such a scenario, we first need to build a coordinate system to express the vertical and horizontal optical emission angles. The inertial navigation system is most commonly used underwater, which can provide accurate location information within a short term underwater. Nonetheless, its accuracy error cumulatively increases with time, which results in long-term underperformance. Hence, it usually requires periodic calibration combined with other navigation systems, \textit{e.g.} GPS. As it is also expensive to be equipped, the inertial navigation system is not suitable to realize the positioning for the UWNs.

To build a coordinate system in three-dimensional space, we need to build the frames of reference. Naturally, we first employ gravity as the vertical reference. For the horizontal reference, the UWNs lack reliable information on latitude and longitude, where the ellipsoidal coordinate system cannot be built. Thus, we utilize the Earth's geomagnetic field as the common horizontal reference of the BS and the UWNs, thanks to the stable geomagnetic field undersea. For instance, we set the due north as the emission degree zero to build polar coordinates. Therefore, the BS can provide the relative angles in vertical and horizontal directions, which are also the emission directions of the optical transmitters.

\subsection{UWN Identification}
\label{sec:UWN-identification}
The identification is based on a three-way handshake mechanism. Section \ref{sec:location} has presented the first handshake request sent from the BS, where the optical emission angles are provided. Then, the UWNs need to match the allocated channels to establish the optical links.

After the BS has assigned the network resources, another question lies in how the UWNs match the allocated channels, emission angles, and IDs. The UWNs cannot respond and apply for access as they lack the emission angles for the LOS communications. Thus, the network identification is unidirectional, which needs a specific feature as the reference for each UWN to match the given resource. The most easy-to-get underwater feature of a UWN is depth, which can be easily and economically measured by both BS and UWN itself. Thus, in the first handshake of the identification process, the downward acoustic message should contain the corresponding depth information of the corresponding UWNs. The successful match is considered as the allocation accomplished on the UWN side. In addition, the UWNs continue receiving the acoustic signal of all the time slots, and they match all the depth information to that of their own.

After the UWNs have matched the depth and obtained the optical emission angles, the UWNs send the optical beam to complete the second handshake. Then, the corresponding acoustic response from the BS completes the third handshake of the identification.

\subsection{Decomposition Algorithm}
\label{sec:decompose}
In realistic scenarios, it is more than probable that the UWNs are deployed with the same depth due to the application requirement. Furthermore, the depth sounding accuracy of the BS is limited, which also depends on the target’s depth, \textit{e.g.} larger depth of the target leads to a lower depth sounding accuracy. These will raise the probability that the BS fails in distinguishing the depth differences of the UWNs, and consequently the downward depth information will be sent with the same values. Therefore, with several same depths, the UWNs may match several IDs. In this subsection, we propose a decomposition algorithm to solve this problem.

First, the BS assigns a marker identifier to mark whether the depth conflicts exist in each time slot. After the acoustic signals are received, each UWN will match its depth with the demodulated information. Some UWNs may find the matched ID exists conflict with other UWN(s) diving in the same depth. These conflicting UWNs are then required to dive or raise with random vertical velocity in a random time, which is also easy to achieve for the underwater terminals. At the same time, they continue receiving the downward acoustic signals from the BS to match the depth. The conflicting UWNs should remain the movement until their depths are distinguished as unique. A marker identifier is also assigned to mark whether the corresponding UWN is diving or rising to accelerate this process. Nonetheless, the vertical velocity differences among the large number of conflicting UWNs have the probability of being too small due to the low mobility of the UWNs, which may result dramatical increase of the decomposition delay. After the timeouts, the decomposition of this round is failed. Then, the BS informs the conflicting UWNs to reset their vertical velocities at each resigned time. The BS also assigns 1 bit in the acoustic frame to control the reset.

After the UWNs have completed the identification, they will return to the former depth according to the requirement of the application. Note that each successful identification can also accelerate the process of the decomposition for other conflicting UWNs.

\begin{figure*}[t]
   \centering
   \includegraphics[width=140mm]{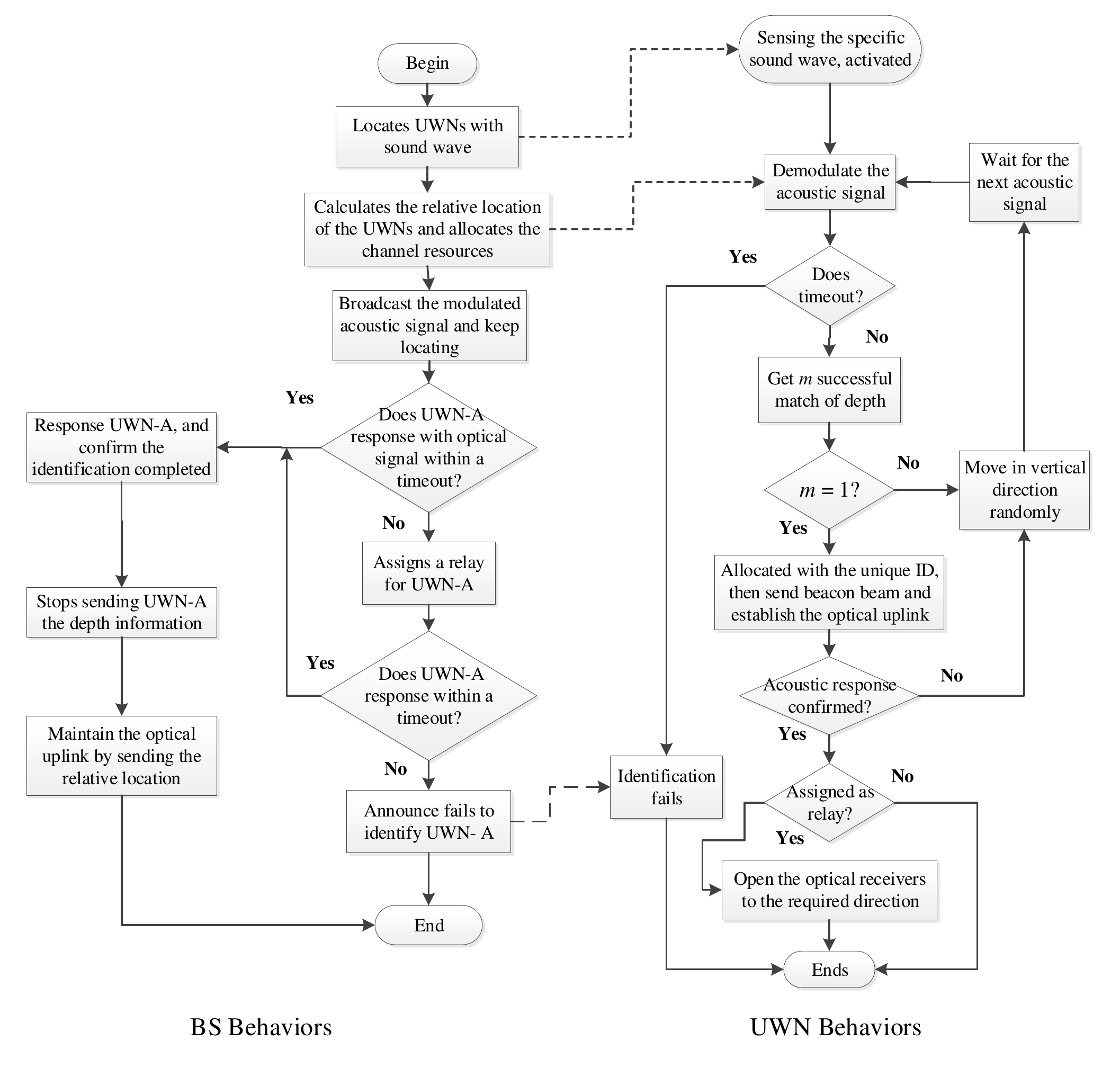}
    \vspace*{-0.5\baselineskip}
    \caption{Flow Charts of BS Behaviors and UWN Behaviors of the proposed method.}
    \label{fig:Flow-Chart}
\end{figure*}

\subsection{Underwater Optical Relaying}
The underwater environment is unfriendly for wireless data transmission. The UWNs may not be able to achieve the LOS communications to the BS via single hop for the following reasons:

\begin{itemize}
    \item The range of the UWAC transmission is much larger than that of the UWOC transmission and the sound propagation shows little dependency on the water quality. Thus, the BS may locate the UWNs that are out of the transmission reach of the UWOC. These UWNs are activated to send optical signal to the BS, while the optical beam cannot be received by the BS, due to the high water attenuation coefficient, \textit{e.g.} turbid harbor water, and strong underwater turbulence.
    \item All the optical or acoustic channels of the BS are assigned and occupied. Thus, the BS will not assign an optical channel for the newly detected UWNs. The UWNs need to wait for the resource assignment.
    \item As the LOS communications, the optical beam may be covered by underwater obstacles, such as fish swarms, seafloor reefs, marine debris, and other UWNs.
    \item Limited by underwater detection and recognition precision, some underwater objects may be misidentified by the BS. However, the BS still assigns certain network resources for them.
    \item Some UWNs do not intend to join the network. For instance, some UWNs are not the target nodes that the BS needs to collect data from, and they have different awaken systems. Thus, they did not receive the corresponding preset sound tag or specific sound frequency in the initialization step, and they remain silent throughout the networking.
    \item As most underwater sensors are of large amount and low cost, the terminal recycling is usually more expensive than the terminal itself. Thus, these nodes are designed and deployed as disposable. After the nodes are out of power, they remain to stay at the same place without recycling. Although they can be correctly detected and recognized, they are not able to respond.
\end{itemize}

Unfortunately, lacking valid and practical means, it is hard to find out the actual reason when the BS does not receive the optical beam from the UWNs. In some cases, \textit{e.g.} the optical beam from the UWNs cannot reach the BS, the UWNs need more time or other methods to access the BS. Thus, the corresponding methods should be explored.

When a UWN has been informed of the unique depth without conflict but not yet establishing the uplink, the BS will repeat the first handshake process before the UWN responds or timeouts. Thus, for the timeout UWN, the BS will assign an accessed node as the relay, usually the closest one to it. The BS then notifies the relay node to prepare to receive the optical beam from the timeout UWN, \textit{i.e.} the relay node adjusts its optical receiver to the relative angles of the UWN. The timeout UWN is also informed of the relative emission angles of the relay node instead of that of the BS. Therefore, the timeout UWN can transfer data to the BS through the dual-hop optical uplinks and receive the acoustic messages directly from the BS via downlink. However, if the timeout UWN still cannot complete the second handshake process, the BS will then announce the identification fails and not assign the network resources for them.

The optical link is severely affected by depth, visibility in the water, and optical emission zenith angle. Thanks to the assigned relay nodes, the optical transmission range can be extended to hundreds of meters \cite{ZZeng2017CST}.

\begin{figure*}[t]
  \centering
  \includegraphics[width=160mm]{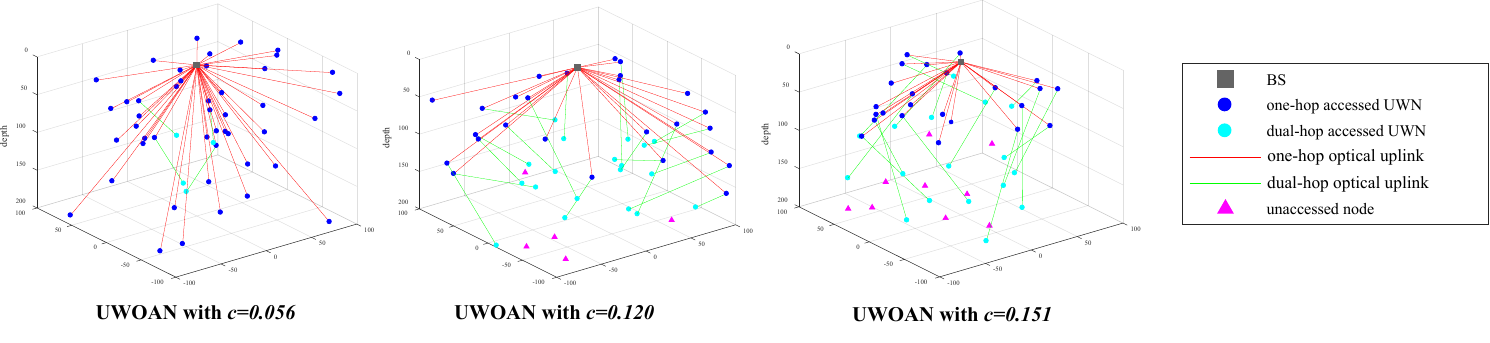}
    \vspace*{-0.5\baselineskip}
    \caption{Randomly generated UWOAN topological graphs of three scenarios with different optical attenuation coefficients $c$ and at most one single relay after networking initialization}.
    \label{fig:UWOAN}
\end{figure*}

\subsection{Network Initialization Summary}
In this subsection, we depict the proposed networking initialization method by using flow charts, which shows the behaviors during the networking for the access-needed UWN and the BS. Figure \ref{fig:Flow-Chart} shows the behaviors of the BS (the left flow chart) to identify the UWN-A and the behavior of UWN-A (the right flow chart). The procedures can be briefly summarized as the behaviors of BS and UWN-A, which are described in the following:

\begin{itemize}
    \item The BS enters into the designated sea area where the UWNs are deployed underwater.
    \item The BS sends the sound waves and receives the echo to detect the UWNs. Note that it is an economical way if the acoustic frequency of the detecting sound wave is the same as the one to activate the UWNs. Alternatively, the detecting sound wave can carry the specific sound tag to activate the UWNs.
    \item The UWNs are activated by the specific sound tag or acoustic frequency. The activated UWNs start to receive and demodulate the acoustic signals.
    \item The BS allocates the channel resources based on the detection. The BS sends the acoustic signal carrying depth information to assign network resources and IDs.
    \item Some UWNs may match to more than 1 IDs since they are in the same depth. Then for these UWNs, they will start the decomposition procedures and stop moving till they get a unique ID or the timeouts.
    \item Once the UWNs get the unique ID, they also obtain the control message from the BS. Thus, they send the optical beam to the given emission direction.
    \item The BS sends the information of access confirmation to the corresponding UWNs, which signifies the accomplishment of the access procedure. Note that the UWNs may not receive the information of access confirmation for various reasons as aforementioned.
    \item The successfully accessed UWNs may be assigned as the relay for the un-accessed UWNs. They need to turn the optical receiver to the given direction.
\end{itemize}

Figure \ref{fig:UWOAN} shows the simulation results of the UWOAN connectivity in three different situations via networking initialization. The simulations are conducted in the inhomogeneous underwater scenario with different optical attenuation coefficients $c$. We assume that 50 UWNs are randomly deployed in a $200$ $m$ $\times$ $200$ $m$ $\times$ $200$ $m$ underwater space nearing surface, and a BS deployed on the water surface. The maximum networking initialization time is set as $50$ s. Seawater is unevenly distributed in the vertical direction, and several underwater parameters, such as temperature, pressure, salinity, and so on, vary with the depth. Thus, the link models are based on the LOS transmission in the inhomogeneous underwater scenario \cite{Zhang2020ITJ}. Other network parameters are detailed in \cite{ZZeng2017CST,Zhang2020ITJ}. The UWNs can access the BS with at most one relay node. With the concerns of the transmission capacity and the LOS UWOC of the UWN, the relay node can be assigned to relay at most one UWN. The deployment of the relay can significantly increase the network coverage area. However, although no blockage of sea obstacle is set, the network coverage area significantly decreases as the ocean water gets turbid. Note that the results are not always the same even with the same coordinates. The vertical random movement of the UWNs in the decomposition algorithm may cause the final topology to differ in each simulation.

\section{Open Challenges and Future Works}
The current research on underwater networking adopt either the costly implementations for underwater terminals or the ideal assumptions for the transmission environment. These proposals cannot be directly applied to realistic underwater scenarios. As we propose a practical underwater network for the UWOAN, there are still open challenges to be solved, and several future explorations to be done. They are summarized as follows.

\begin{table}[t]
\centering 
\caption{Access Quality with 10,000 Repeated Random Simulations Based on Three Underwater Scenarios.}
\label{tab:access}
\scalebox{0.48}{
\resizebox{\textwidth}{!}{
\begin{tabular}{cccc}
\toprule
Optical attenuation coefficient $c$ & $0.056$ & $0.120$ & $0.151$\\
\toprule
Average sound propagation delay (s) & $0.6540$ & $0.6419$ & $0.4693$\\
\midrule
Maximum decomposition delay (s) & $9.5398$ & $9.7771$ & $9.7804$\\
\midrule
Access rate & $99.58\%$ & $87.93\%$ & $68.77\%$\\
\midrule
Dual-hop access rate & $8.24\%$ & $33.65\%$ & $28.01\%$\\
\bottomrule
\end{tabular}
}
}

\vspace{2mm}
-With 50 UWNs randomly deployed in a $200$ $m$ $\times$ $200$ $m$ $\times$ $200$ $m$ underwater space nearing surface.\\
\end{table}

\subsection{UWOC Link Maintenance}
The transmission range of the UWOC severely limits the network coverage of the UWOAN. The optical relay deployment is a feasible way to extend the scalability of the networks. As shown in Table \ref{tab:access}, the access rate can be raised by up to $33.65\%$ via optical relay assignment. Theoretically, the cover area of UWOAN can be extended by the multiple-hop optical links. However, two challenges are critical to be faced. 1) The UWOC link severely suffers from misalignment, and it is especially serious for the mobile nodes. Meanwhile, the underwater currents also lead to the nodes quiver passively, and the misalignment still exists for the anchored UWNs. In FSO, the satellites use a spin mechanism to overcome this phenomenon. Such methods need to be explored for UWOC. 2) As an LOS communication, UWOC can also be blocked by sea lives for a relatively long time. The LEDs with wide divergence angles can relax the requirement of LOS. However, it comes with the price of a drastically reduced transmission range. As the consequence, the UWOC path with multiple hops is unreliable. To maintain the optical links, the downward acoustic signal can continuously provide the information of real-time angles to support the UWNs tracking the optical receiver. The corresponding researches for multiple relay access schemes need to be explored.

\subsection{Delay}
For the long underwater path in the UWOAN, the delay becomes the main challenge. It includes two aspects: decomposition delay and transmission delay. The decomposition delay is mainly caused when two or more UWNs diving in the same depth. These UWNs need to create enough depth difference by randomly diving, raising, and keeping stationary to be distinguished. The delay is related to their speeds, the moving directions, and the depth. The Table \ref{tab:access} shows the decomposition delay is near $10$ $s$. An efficient depth adjustment method would significantly reduce the decomposition delay, as well as the energy consumption. Alternatively, the game theory can be used to solve the problem when UWNs are in such a prisoner's dilemma. Table \ref{tab:access} also shows the sound propagation delay is up to $0.65$ $s$ with a range of $200$ $m$. Such a delay may cause the alignment information from the BS postponed. Consequently, the UWNs would suffer from severe optical misalignment in the environment with underwater turbulence, which is unacceptable for real-time applications. 

\subsection{BS Deployment}
As the transmission range is strongly limited by the water attenuation, the optimal BS deployment can increase the coverage area for the UWOAN. For instance, the BS deployed on the edge of the UWN swarm may lose access to UWNs on the other side of the swarm. For a large UWN swarm, several BSs need to be assigned. The locations of these BSs should be optimized to cover all the UWNs. For such a scenario, a UWN may receive more than one sound signal at the initialization. It may cause the UWN matching the erroneous channel and failing in access. Thus, the corresponding multiple access strategies need to be investigated.

% \subsection{Interoperability}
% The UWOAN is essentially a heterogeneous network consisting of various devices and different communication modes. Interoperability is more than important for such a mixed system. It consists of two aspects, interoperability for communication techniques and interoperability for devices. The communication modes, apart from the UWOC and UWAC as we have previously discussed in the UWOAN, also include the data transmission modes from the BS to the on-shore data center, \textit{e.g.} the wired transmission via submarine cable, the wireless communication to the on-shore access points, and the satellite transmission. The marine devices have an even more complex variety, \textit{e.g.} ships, submarines, buoys, sensors, AUVs, UAVs, etc. Hence, the interoperability for the UWOAN first needs a scalable framework to standardize the development of the UWOAN, then a series and a large number of standards for these communication modes and marine devices.

\section{Conclusions}
In this article, we proposed a novel network initialization method for hybrid UWOANs. Due to the lack of GPS service and limited device payload in underwater environments, we make use of acoustic downlinks for UWN localization and signaling. In light of this, the proposed method decomposes the network initialization procedure into multiple components including UWN identification, discovery, localization and conflict decomposition. Numerical results demonstrate the effectiveness of the proposed solution. Besides, some open challenges in UWOANs are also discussed. 

% \section{Acknowledgment}
% The work is partially supported by Eiffel Excellence Scholarship (No.P745849E), China Scholarship Council (202006960046), Campus France PHC Cai Yuanpei 2019 project (44016XA), the 863 High Technology Plan of China (2013AA013402), the National Natural Science Foundation of China (61172080, 61771357), and the open project (2020GZKF017) of the State Key Laboratory of Advanced Optical Communication Systems and Networks, Shanghai Jiao Tong University, China.

\bibliographystyle{IEEEtran}
\bibliography{biblio}

% \begin{IEEEbiography}
% \section{Biography}
\vspace{-1.2cm}
\begin{IEEEbiographynophoto}{Yuanhao Liu}
is currently a Ph.D. student with the State Key Laboratory Integrated Service Networks, School of Telecommunications Engineering, Xidian University, China, where he received a B.E. degree in 2016. He is also a Ph.D. student with CERI-LIA, Computer Science Laboratory, University of Avignon, France, and IMT Lille Douai, Institut Mines-T\'el\'ecom, Univ. Lille, Center for Digital Systems, France. His research interests include underwater optical communications, and protection in optical networks.
\end{IEEEbiographynophoto}
\vspace{-1.5cm}
\begin{IEEEbiographynophoto}{Fen Zhou}
(SM’15) received the Ph.D. degree in networking from INSA of Rennes, France in 2010. He is currently a full professor at IMT Nord Europe, Institut Mines-T\'el\'ecom, Univ. Lille, Center for Digital Systems, France. His research interests include network security and survivability, routing and resource allocation, network function virtualization, as well as routing optimization in intelligent transportation systems. %He has served as the Symposium Co-Chair, the Publicity Co-Chair, the Local Organization Chair, and the Session Chair in several international conferences, such as the IEEE ICNC18, IEEE WCSP14, IEEE Wimob15, NetgCOOP16, IEEE MovNet17, and IEEE Globecom13. He has also served on the program committees for several international conferences such as INFOCOM, ICC, GLOBECOM, and ONDM.
\end{IEEEbiographynophoto}
\vspace{-1.5cm}
\begin{IEEEbiographynophoto}{Tao Shang}
received his B.E. and M.E. degrees from Huazhong University of Science and Technology, Wuhan, China, in 1994 and 2001, respectively, and his Ph.D. degree from Shanghai Jiao Tong University, Shanghai, China, in 2006. Now, he is a full professor at the State Key Laboratory of Integrated Service Network, School of Telecommunications Engineering, Xidian University, Xian, China. His main research topic covers photonic devices and subsystems, optical networking, and wireless laser communication.
\end{IEEEbiographynophoto}

\end{document}